\input psfig.sty
\input epsf.sty

\def\mev{\,{\rm Me\kern-0.1em V}}
\def\gev{\,{\rm Ge\kern-0.1em V}}

\documentstyle[12pt]{article}

\begin{document}
\begin{center}
{\Large{\bf  Unquenched QCD with Light Quarks}}\\
\vspace*{.45in}
{\large{A. ~Duncan$^1$, E. ~Eichten$^2$,
and J. ~Yoo$^1$}} \\
\vspace*{.15in}
$^1$Dept. of Physics and Astronomy, Univ. of Pittsburgh, 
Pittsburgh, PA 15260\\
$^2$Theory Group, Fermilab, PO Box 500, Batavia, IL 60510\\
\end{center}
\vspace*{.3in}
\begin{abstract}
We present recent results in unquenched lattice QCD with two degenerate light
sea quarks using the truncated determinant approximation (TDA).
In the TDA the infrared modes contributing to the quark determinant are computed 
exactly up to some cutoff in quark off-shellness (typically 2$\Lambda_{QCD}$).
This approach allows simulations to be performed at much lighter 
quark masses than possible with conventional hybrid MonteCarlo techniques.
Results for the static energy and topological charge
distributions are presented using a large ensemble generated 
on very coarse (6$^4$) but physically large lattices. 
Preliminary results are also reported for the static energy 
and meson spectrum on 10$^3$x20 lattices (lattice scale $a^{-1}$=1.15 GeV) 
at quark masses corresponding to pions of mass $\leq$ 200 MeV.  
Using multiboson simulation to compute the ultraviolet part of the quark determinant 
the TDA approach becomes an exact algorithm with essentially no increase in computational effort.
Some preliminary results using
this fully unquenched algorithm are presented.
\end{abstract}
\newpage

\section{Introduction}

Although much progress has been made in the last few years towards the goal
of simulating important hadronic quantities in fully unquenched lattice QCD,
the physical regime of light up and down sea quarks (quark masses $<$10 MeV)
remains basically intractable with the current standard algorithms (hybrid
MonteCarlo (HMC) and variants thereof \cite{HMC}) even with the Teraflop 
scale parallel platforms which are presently coming online. With these algorithms, 
the generation of a statistically significant ensemble of dynamical gauge 
configurations with ``light" quark masses chosen to give a pion mass just 
light enough to allow rho decay (on sufficiently large lattices) already 
consumes many Teraflop years of computational effort \cite{Lippert}. The high 
cost  of performing unquenched HMC simulations with light sea quarks arises 
from two independent sources: the sensitivity of conjugate gradient solvers of 
large systems to the condition number of the matrix to be inverted, and the 
growing autocorrelation time due to critical slowing down as the critical 
point corresponding to the chiral limit at which the pseudoscalar mass 
vanishes is approached. Most of the power growth (as a function of inverse 
pion mass) in the computational effort is in fact due to the first of these 
sources \cite{Lippert}, which the algorithms described in this
paper are designed to circumvent. 

Let us first recall the origin of the problems with the iterative solvers
used in typical HMC simulations of unquenched QCD. To be specific, we
consider throughout the case of Wilson (or Sheikoleslami-Wohlert ``clover"
improved) quark actions. The desired quark
determinant (with two degenerate flavors of sea quark) is introduced via
pseudofermionic fields with a quadratic action involving the {\em inverse}
of the squared quark (Wilson/clover)-Dirac operator $Q$. The system is
then treated as a classical Hamiltonian one subjected to molecular dynamics
evolution corresponding to the following Hamiltonian 
\begin{eqnarray}
\label{eq:HMC}
     \cal{H}_{MD} &=& \frac{1}{2}\rm{Tr}(P_{U}^{2})+S_{\rm gauge}(U)+\phi^{\dagger}(Q^{\dagger}Q)^{-1}\phi \\
      Q^{\dagger}Q &=& H^{2},\;\;\;\; H\equiv \gamma_{5}Q \;\;\;(\rm{hermitian})
\end{eqnarray}
where $P_{U}$ are the conjugate momenta to the gauge-fields $U$, $S_{\rm gauge}(U)$ is the
pure gauge action, and $\phi$ is a bosonic field with a highly nonlocal action. In order
to update the gauge-fields, the force on these fields due to the $\phi$ field must be
computed, and this involves the inversion of the $Q^{\dagger}Q$ operator, which is identical
to the square of the {\em hermitian} Wilson-Dirac operator $H\equiv \gamma_{5}Q$. As the
quark mass is taken to zero, the operator $H$ frequently develops very small eigenvalues:
equivalently, the condition number (ratio of highest to lowest eigenvalue) not uncommonly
becomes very large (in simulations described later in this paper, condition numbers $>$1000
are quite common for $H$). The required inversion of $H^{2}$ in Eq(\ref{eq:HMC}) then involves
an operator of condition number $>$10$^{6}$, which not surprisingly requires a very large
number of conjugate gradient sweeps. In a nutshell, most of the computational difficulty
with very light sea-quarks in the standard hybrid MonteCarlo algorithm arises from this source. 

In contrast to the ``freezing" problem encountered with linear solvers of the 
conjugate gradient variety, the  extraction of low eigenvalues 
by Krylov subspace methods such as the Lanczos algorithm \cite{Lanczos} 
does {\em not} deteriorate as a consequence of the presence
of a very small eigenvalue. The rapidity with which the Lanczos procedure extracts eigenvalues
in a given region of the spectrum is instead determined by the local spectral density in
that region, which for the operator $H$ of interest to us here is in fact minimal near zero.
The eigenvalues of $H$ (cf. Section 2) have the physical interpretation as a gauge-invariant
extension of quark off-shellness in the free theory, so the truncation of the full quark 
determinant to a product of all modes with (absolute value) eigenvalue below some cutoff $\mu$
corresponds to a gauge-invariant approximation of the fully unquenched theory in which
sea-quark loops up to quark off-shellness $\mu$ are  included exactly and completely. This
approximation will be referred to as the ``truncated determinant approximation" (TDA)
in the following. In previous publications various features of the implementation of this
algorithm have been discussed \cite{TDA1,TDA2}, as well as the application to the study
of stringbreaking on large coarse lattices \cite{stringbreaking}. 
The work described here is motivated to a large extent by a desire to provide 
alternatives to HMC which would allow at least some quantities to be computed directly 
in the deep chiral regime as a check on the large extrapolations required from the 
quark masses presently practicable in the HMC approach to the physical range.

In Section 2, we review
the basic features of a truncated determinant approach to unquenched QCD in the light of
the much more extensive simulations which we have performed since the aforementioned
references. In particular, we argue that many low-energy hadronic quantities (e.g. low-lying
hadron spectrum, string-breaking, low energy chiral physics) can be studied quite precisely
in the TDA approximation, while other effects which depend more sensitively on the ultraviolet
structure of internal quark loops (e.g. channels involving the eta-prime) require a fully
consistent treatment of the full quark determinant. 
In Section 3 we present recent results obtained
with large ensembles ($\simeq$10000 configurations) of unquenched TDA configurations on physically
large, coarse lattices ($6^4$, with O($a^2$) gauge action improvement): the quantities studied 
include the static energy of a heavy quark-antiquark pair, and dependence of topological charge
distributions on the quark mass. In Section 4 we present some preliminary results obtained on larger
lattices (10$^3$x20, with a lattice scale $a^{-1}$=1.15 GeV): here we concentrate on extracting 
the low-lying meson spectrum at up and down sea-quark masses close to 
their physical values ($m_{\pi}/m_{\rho} < 0.26$). We also give an
example of a correlator in which the truncated determinant approximation introduces a
visible anomaly analogous (but quantitatively less severe) to the one familiar from 
quenched calculations: namely, the scalar isovector channel \cite{scpaper}.  Finally, in Section 5, we describe
a combined TDA + multiboson approach which allows exact unquenched simulations in the very light
quark regime with a small number of multiboson fields. Some preliminary results of simulations on
large coarse lattices with this technique are described.

\newpage
\section{Infrared and Ultraviolet Quark Modes: the Truncated Determinant Approximation}

The hermitian (Euclidean) Dirac operator $H \equiv \gamma_{5}(D\!\!\!\!/(A)-m)$ has a spectrum which can
be regarded as the gauge-invariant generalization to nontrivial gauge fields of the quark off-shellness
of the free quark theory. 
Indeed, the eigenvalues of the free operator $H_{0} \equiv \gamma_{5}(D\!\!\!\!/(A=0)-m)$ are
just $\pm \sqrt{p^{2}+m^{2}}$, which precisely corresponds to the signed Euclidean off-shellness of
a quark of mass $m$ and momentum $p$. Moreover, the individual eigenvalues (though not, of course, the
eigenvectors) are gauge-invariant. Roughly speaking, we can therefore visualize the contribution to the
quark determinant from the infrared modes (corresponding to the eigenvalues of $H$ of smallest absolute
value) as arising from quark loops of large physical extension in Euclidean coordinate space, while the
ultraviolet modes correspond to quark loops of small size. Gauge-invariant quark loops of small size
correspond to the lowest dimension gauge-invariant operators so we should expect that the contribution
to the quark determinant from the highest UV modes amounts to a functional of exactly the same form
(i.e. $\int F_{\mu\nu}^{2}d^{4}x$ in the continuum) as the basic pure gauge action, and therefore has
the  sole physical effect of changing the scale  in any gauge-invariantly cutoff version of the theory,
 such as lattice QCD.  
 
  To make these arguments a little more concrete, let us imagine separating low and high quark eigenmodes
in an analytically smooth way by switching off the higher eigenvalues above a sliding scale $\mu$. 
If we define
\begin{equation}
  D(\mu) \equiv \frac{1}{2} \rm{tr\; ln}(\tanh{\frac{H^{2}}{\mu^{2}}})
\end{equation}
then a weak-coupling expansion of $D(\mu)$ shows \cite{TDA1} that the $\mu$ dependence is given asymptotically
for large $\mu$  by
\begin{equation}
  D(\mu)\simeq \beta_{F}\ln{\frac{\mu^{2}}{m_{q}^2}}\int d^{4}x F_{\mu\nu}^2 + O(\frac{1}{\mu^{2}}(DF)^{2})
\end{equation}
where $\beta_{F}$ is the one-loop quark contribution to the beta function. This result illustrates in an
explicit way the role of the high quark modes in renormalizing the pure gauge action, which physically
corresponds to the screening of the gauge interactions by virtual quark-antiquark pairs.

  The above considerations suggest in the case of lattice QCD a natural (and gauge-invariant) truncation
of the full theory in which the quark determinant is split into an infrared and ultraviolet part at 
an appropriately chosen scale $\mu$. In the lattice case, the operator $H$ is a large, sparse hermitian matrix
with eigenvalues $\lambda_{i}$, and we may write
\begin{equation} 
   \rm{det}(H^2) = (\prod_{|\lambda_i|<\mu} \lambda_{i}^{2})(\prod_{|\lambda_{i}|>\mu}
               \lambda_{i}^{2})\equiv D_{IR}(\mu)\cdot D_{UV}(\mu)
\end{equation}  
for the quark determinant appropriate for two degenerate sea-quark flavors. For sufficiently high $\mu$, the
contribution of $D_{UV}$ to the effective action (after integrating out quark fields) should amount to a
renormalization of the pure gauge action. Specifically, we expect that $D_{UV}$ should be accurately modelled
by small Wilson loops which can then be absorbed into the pure gauge part of the action and should induce, at least
for low energy quantities dominated by processes in which virtual quark off-shellness is typically lower 
than $\mu$, only a change of lattice scale, while leaving dimensionless quantities (such as ratios of hadron
masses) unchanged. This expectation has been confirmed by extensive numerical studies \cite{TDA2}, which show that
in many cases $D_{UV}$ is accurately modelled by linear combinations of loop operators containing 6 links or less. 
In these studies, it was important to perform the IR-UV split by choosing a fixed number $N_{\rm eig}$ of low
eigenvalues: $\mu$ is then the average value of $|\lambda_{N_{\rm eig}}|$ (where the eigenvalues of $H$ are always 
ordered with respect to absolute magnitude). Otherwise, a small variation of the gauge configuration can cause
$|\lambda_{N_{\rm eig}}|$ to cross the scale $\mu$, resulting in a discontinuous jump in $D_{UV}$, a situation clearly
incompatible with a smooth analytical fit of $D_{UV}$ to a fixed linear combination of small Wilson loops. 
Accordingly, for the rest of this paper, we shall define
\begin{eqnarray}
  D_{IR}(\mu) &\equiv& \ln{\prod_{i=1}^{i=N_{\rm eig}} \lambda_{i}^{2}} \\
  D_{UV}(\mu) &\equiv& \ln{\prod_{i=N_{\rm eig}+1} \lambda_{i}^{2}} \\
  \mu &\equiv& <|\lambda_{N_{\rm eig}}|>
\end{eqnarray}
The truncated determinant approximation (TDA) will correspond to the interpolation between quenched and
full QCD induced by replacing the full quark determinant by the infrared piece $D_{IR}$ in the effective pure
gauge action obtaining after quark fields are integrated out. We shall argue below that a choice of 
truncation scale $\mu \simeq 2\Lambda_{QCD}$ is adequate to preserve both the important low energy chiral
physics of QCD, as well as the low-lying hadron spectrum (lowest states in each channel).
Moreover, the computational difficulty of extracting the (typically, several hundred) eigenvalues needed
for $D_{IR}$ {\em does not} increase as the quark mass is taken to zero on a fixed size lattice, in contrast
to conventional HMC algorithms. 

\section{TDA simulations on large, coarse lattices}

 The TDA approach has been applied previously to a study of stringbreaking in physically large
(2.4 fermi$^4$) lattices at a single value of the sea-quark mass \cite{stringbreaking}. These
simulations have been extended to four different sea-quark masses (with 2 degenerate flavors of
sea-quark) with ensembles 2-2.5 times larger than previously studied. 
In order to study long distance features of the full theory, we work on coarse 6$^4$ lattices
(lattice spacing $a=$0.4 F) but with O($a^2$) improved
gauge action. Following Alford et al \cite{alfordetal}, we improve the gauge action with
a single additional operator, with coefficients tuned to optimize rotational invariance
of the string tension
\begin{eqnarray}
  S(U) &=& \beta_{\rm plaq}\sum_{\rm plaq}\frac{1}{3}{\rm ReTr}(1-U_{\rm plaq}) \nonumber
\\
            &+& \beta_{\rm trt}\sum_{\rm trt}\frac{1}{3}{\rm ReTr}(1-U_{\rm trt})
\end{eqnarray}
where ``trt"  refers to a 8 link loop of generic structure (+x,+y,+x,-y,-x,+y,-x,-y) 
(the ``twisted rectangle" of Ref\cite{alfordetal}).
With the choices $\beta_{\rm plaq}$=3.7,
$\beta_{\rm trt}$=1.04$\beta_{\rm plaq}$,  the violations of rotational invariance
expected on such a coarse lattice are almost completely eliminated so that the static
quark potential becomes a smooth function of lattice radial separation \cite{alfordetal}.
As the quark action is not improved, the lattice spacing quoted here is determined by
matching the initial linear rise of the string tension to a physical value (rather than
by using the rho mass, for example). 

\begin{figure}
\psfig{figure=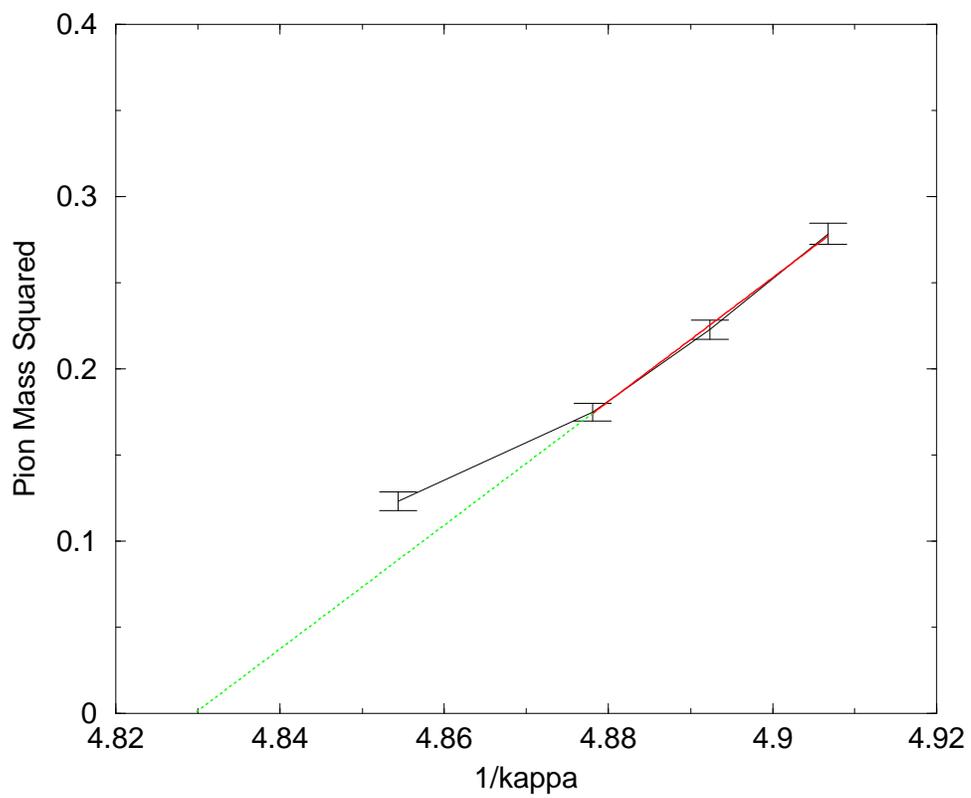,height=0.7\hsize}
\caption{Fit of $M_{\pi}^{2}$ versus $1/\kappa$}
\end{figure}   

  All four values for the sea-quark hopping parameter studied here correspond to very light
quarks by the usual standards of unquenched QCD. In Fig. 1 we show the plot of pion
mass squared versus $\frac{1}{\kappa}$. The lightest quark shows a clear finite volume
effect in the pion mass, so we have determined the critical kappa value from a fit
to the heaviest three quarks only, as indicated in the figure. This fit gives $\kappa_{c}=$0.20706.
In physical units, the three heaviest quarks correspond to pion masses of 210, 235 and 264 MeV,
while the lightest sea-quark studied corresponds to a pion mass of 175 MeV in the finite volume
system, or an infinite volume pion mass of about 150 MeV, very close to the physical value.

\begin{figure}
\psfig{figure=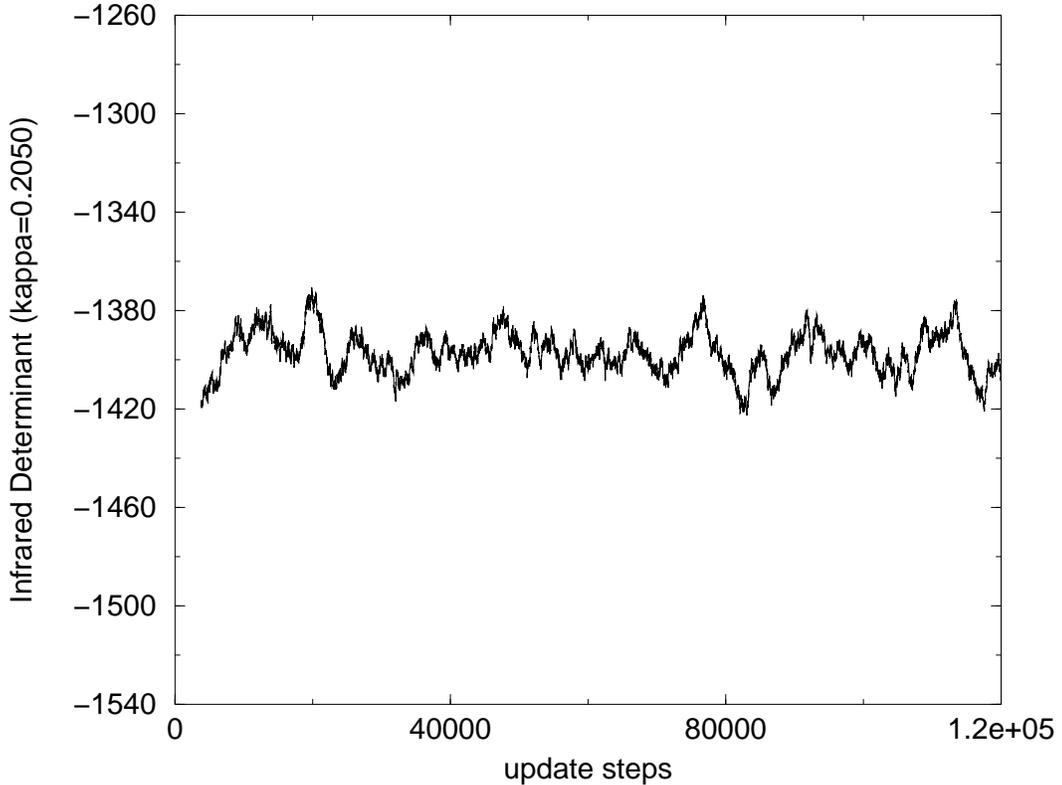,height=0.7\hsize}
\caption{Monte Carlo sequence of $D_{IR}$ for $\kappa$=0.2050}
\end{figure}

\begin{figure}
\psfig{figure=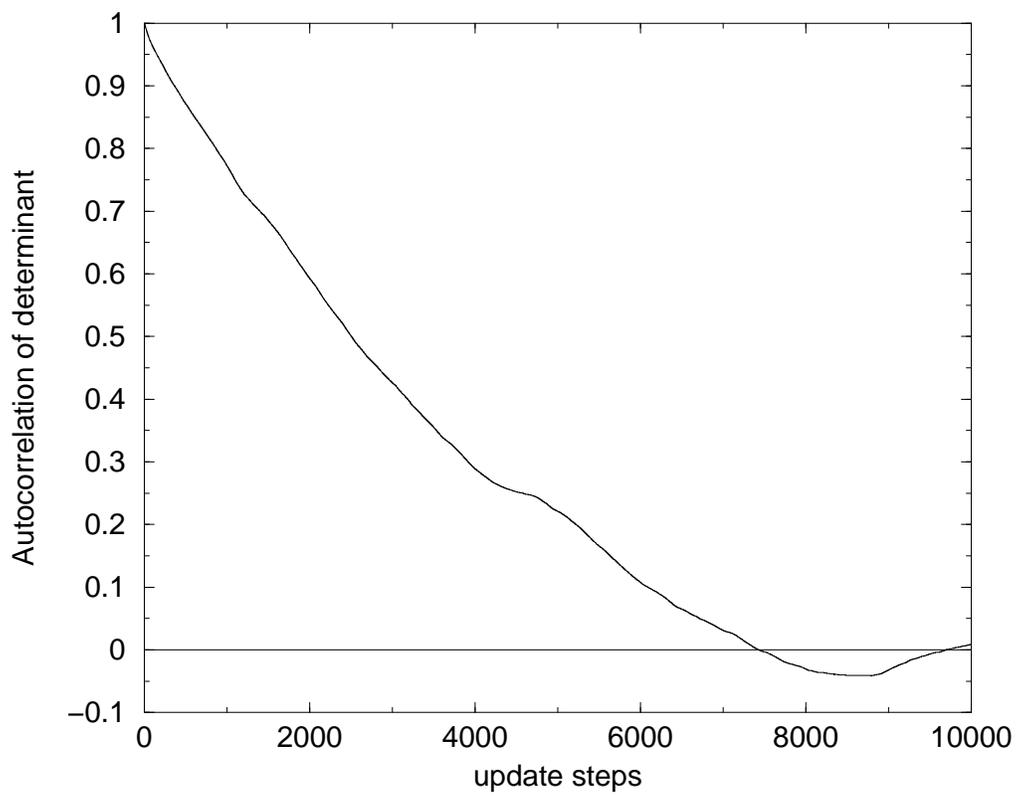,height=0.7\hsize}
\caption{Autocorrelation curve of $D_{IR}$ for $\kappa$=0.2060}
\end{figure}

  In the TDA simulations described here the number of infrared modes included exactly in
the low-energy determinant $D_{IR}$ has been chosen to be 840, for all four kappa values.
The lattice scale as determined from the initial linear rise of the static energy is
essentially unchanged over the limited range of quark masses studied 
 and so is the magnitude of the largest included eigenvalue of $H$
(in lattice units: see Table \ref{tab:lat}), so the scale $\mu$ for the determinant truncation corresponds
in all cases to a physical
off-shellness of approximately $410$ MeV. The global gauge-field update which precedes the accept/reject
step based on the change in $D_{IR}$ is a standard multihit metropolis \cite{stringbreaking},
with parameters chosen to ensure an acceptance ratio on 
the order of 50\% (see Table \ref{tab:lat}).
These parameters are the {\em same} for all four runs, but the acceptance ratio varies only
from 49\% to 58\% even though the quark mass varies by a factor of 3.

  In Table \ref{tab:lat}, we also show the
results of autocorrelation studies of the infrared determinant $D_{IR}$, and of the topological
charge. The sequence of infrared determinant $D_{IR}$ values shows the existence of very long
correlations in this quantity, as is apparent in Fig. 2,   typically extending over thousands
of update steps (1 update step $\equiv$ 1 global gauge-field update followed by an accept/reject
based on $D_{IR}$). This makes it difficult to extract an accurate autocorrelation time, even
with a sequence of 100000 steps. The determinant autocorrelation times $\tau_{\rm det}$ shown
in Table \ref{tab:lat} are obtained by integrating the autocorrelation curves out to a Monte Carlo time
where they first cross zero, but these curves are not even approximately exponential (see Fig. 3), so there
are undoubtedly several important time scales present in the MonteCarlo dynamics for this 
quantity.

\begin{figure}
\psfig{figure=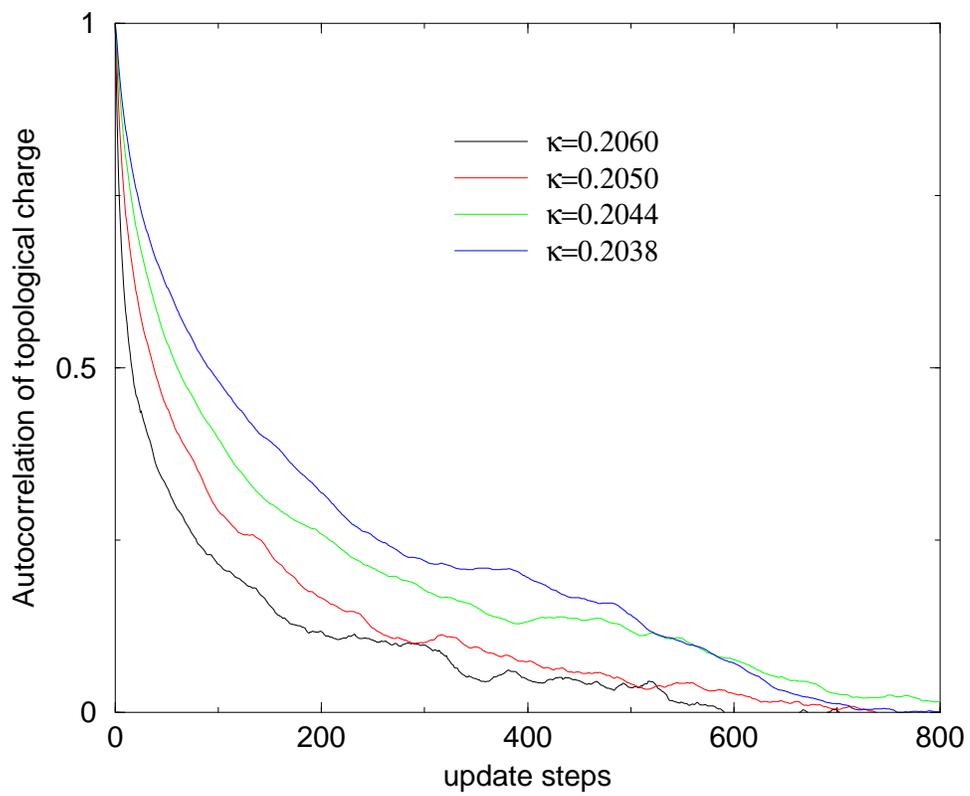,height=0.7\hsize}
\caption{Autocorrelation curves of $Q$ for $\kappa$=0.2060,0.2050,0.2044,0.2038}
\end{figure}

  For the topological charge, the situation is much cleaner.
The topological charge, $Q$, can be expressed \cite{TDA1}
in terms of the eigenvalues of the Wilson-Dirac operator:
\begin{eqnarray*}
 Q = {1 \over 2\kappa} (1 - {\kappa \over\kappa_{c}}) \sum_{i=1}^{N} {1 \over \lambda_i}
\end{eqnarray*}
In practice, this sum is quickly saturated by the low eigenvalues: in particular, we have 
evaluated it by setting $N=N_{\rm eig}$, as these eigenvalues are in any case byproducts
of the TDA update procedure. The autocorrelation curves for $Q$ for the 4 different $\kappa$
values are shown in Fig. 4 and are roughly exponential: the autocorrelation times $\tau_{\rm top}$ given by
the integral of the autocorrelation function are displayed in Table \ref{tab:lat}. For the lighter quarks,
the autocorrelation time determined by an exponential fit at small times is somewhat smaller,
indicating the presence of a longer range component.

\begin{table}
\centering
\caption[lattices]{Characteristics of 4 6$^4$ TDA ensembles, a=0.4 F}     
\label{tab:lat}
\begin{tabular}{|c|c|c|c|c|c|c|c|}
\hline
\hline
Run & $\kappa$  & $M_{\pi}$ & $F_{\pi}$  &$<|\lambda_{N_{\rm eig}}|>$ & accpt. ratio &  $\tau_{\rm det}$ & $\tau_{\rm top}$ \\ 
\hline
\hline
 j1 & 0.2060 & 0.351 $\pm$ 0.008 & 0.19 & 0.826 & 0.49 & $\simeq$2.9x10$^3$  & 75  \\ 
\hline
 j2 & 0.2050 & 0.418 $\pm$ 0.006 & 0.21 & 0.828 & 0.51 & $\simeq$1.4x10$^3$  & 105  \\ 
\hline
 j3 & 0.2044 & 0.472 $\pm$ 0.006 & 0.20 & 0.827 & 0.57 & $\simeq$3.5x10$^3$  & 157  \\ 
\hline
 j4 & 0.2038 & 0.528 $\pm$ 0.006 & 0.18 & 0.827 & 0.57 & $\simeq$2.3x10$^3$  & 182  \\ 
\hline
\hline
\end{tabular}
\end{table}

  The static energy of a heavy quark-antiquark pair has been studied for the four ensembles
described above. Coulomb gauge Wilson lines \cite{stringbreaking} were accumulated after every
update step until a bin size of 2000 steps was reached. The corresponding binned Wilson line
averages (typically, on the order of 40 to 50) were then subjected to a standard bootstrap
analysis, allowing us to extract asymmetric errors. Also, the bin size was varied until the
errors were stable to ensure that autocorrelation effects were eliminated. 
For the lightest two quark masses (runs j1,j2) there is reasonably clear evidence 
of stringbreaking once the Wilson line
ratios are taken between Euclidean times 1.2 fm and 1.6 fm (T=3/4 plots on Figs. 5,6, with the 
large R value agreeing with (twice) the measured mass for a heavy-light meson. With the heaviest
mass sea-quark the levelling off of the static energy at larger distance is less clear (Fig. 7).
Even with the large ensembles collected, it is clear that the sea-quark shielding of the
string tension induces very large fluctuations which makes high precision very hard to achieve.

\begin{figure}
\psfig{figure=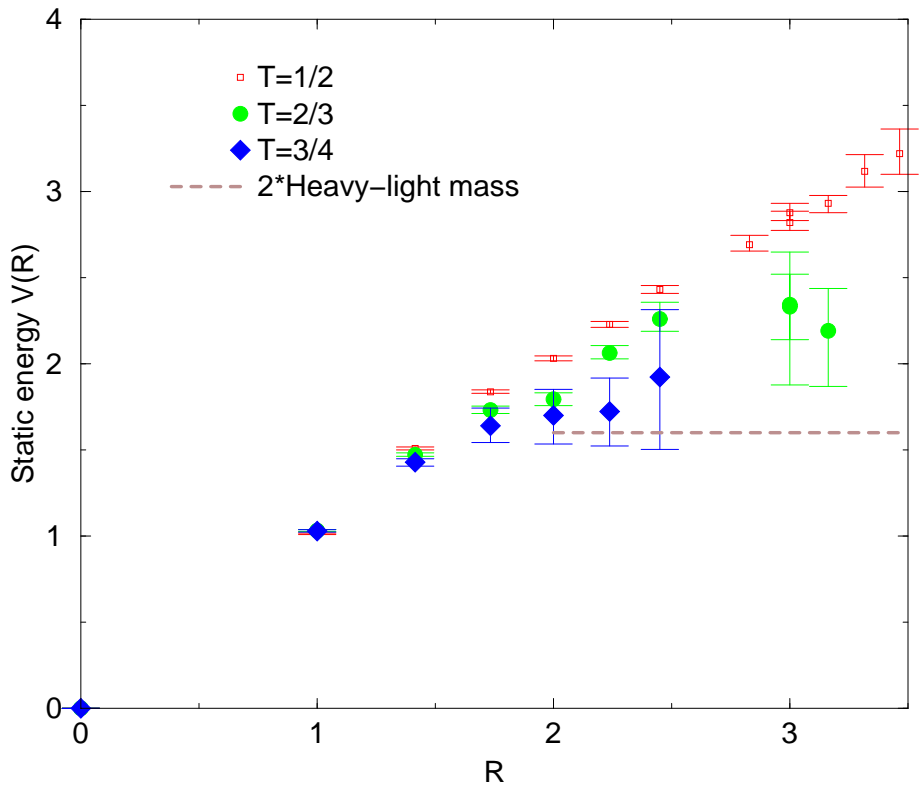,height=0.7\hsize}
\caption{Static Energy $V(R)$ for $\kappa$=0.2060 (j1 run)}
\end{figure}

\begin{figure}
\psfig{figure=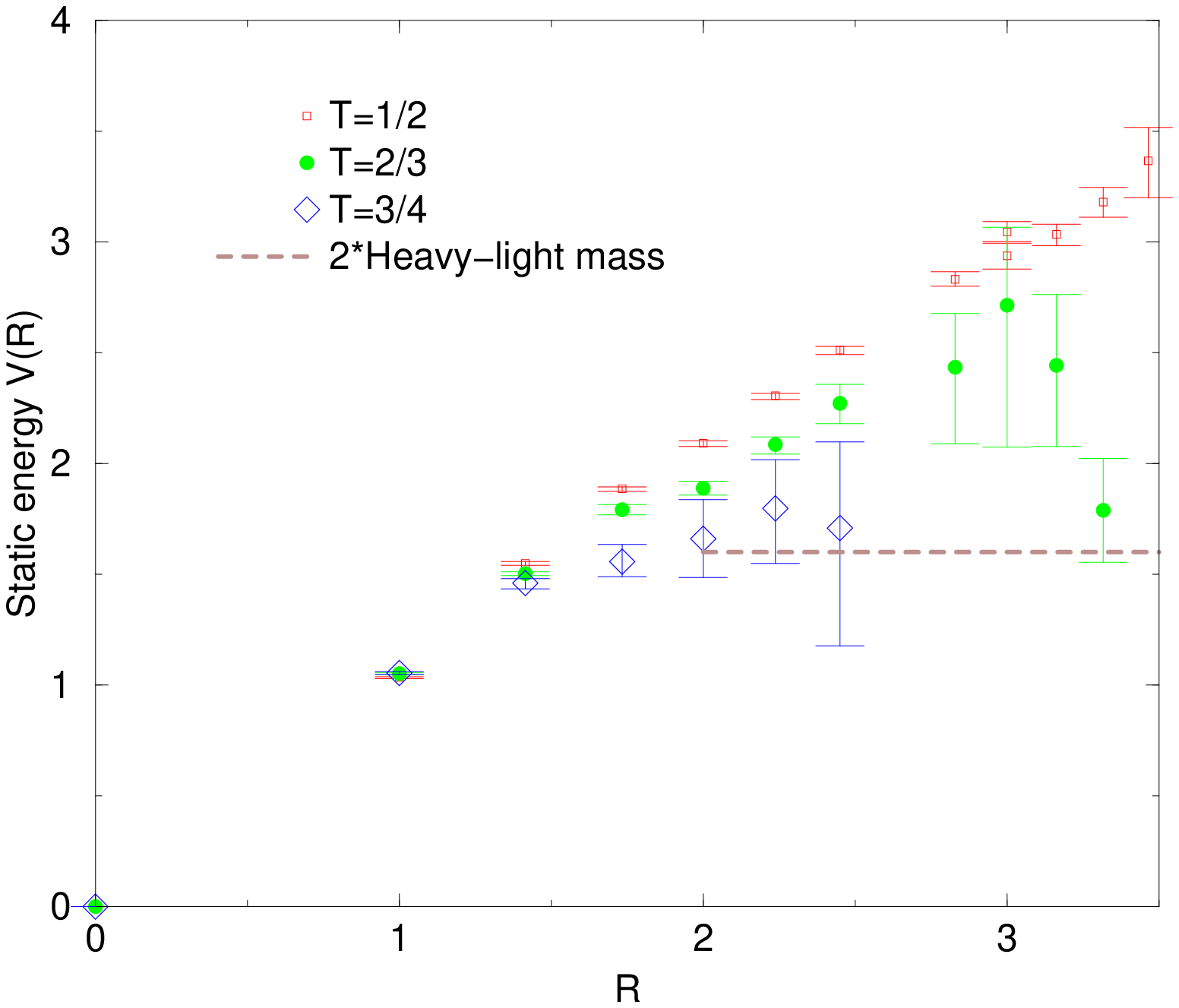,height=0.7\hsize}
\caption{Static Energy $V(R)$ for $\kappa$=0.2050 (j2 run)}
\end{figure}

\begin{figure}
\psfig{figure=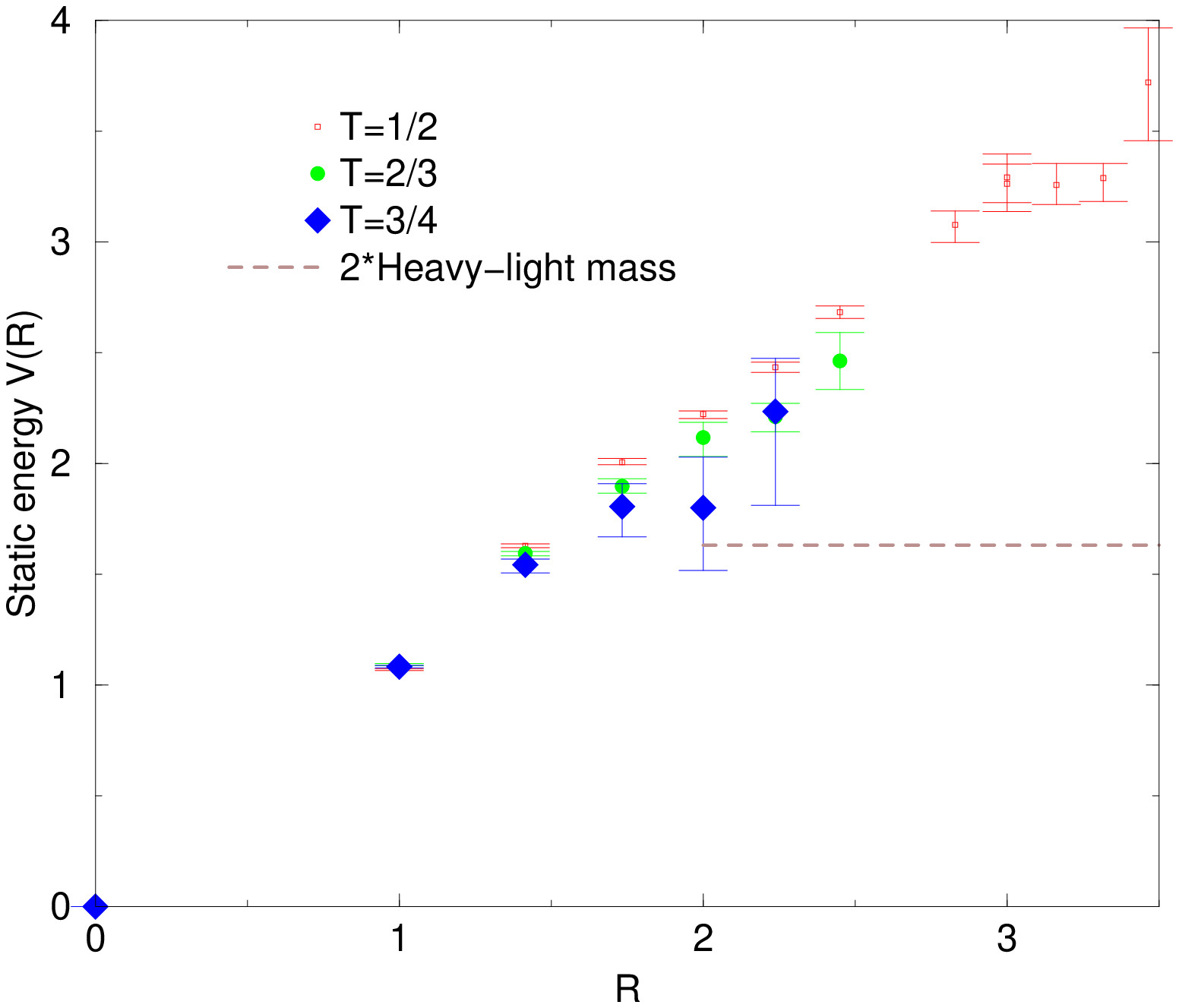,height=0.7\hsize}
\caption{Static Energy $V(R)$ for $\kappa$=0.2038 (j4 run)}
\end{figure}

  Another very characteristic feature of unquenched QCD in the chiral limit arises
from the suppression of nontrivial topological charge as the quark mass goes to zero.
The distribution of topological charge $Q$ is known \cite{LeutSmilg} to follow directly from
the chiral symmetry of QCD. In the case of a theory with two degenerate light quark
flavors, the normalized probability distribution of $Q$ in a system of finite 
space-time volume $V$ is given by
\begin{eqnarray}
\label{eq:topformula}
   P(Q) &=& x\frac{I_{Q}(x)^{2}-I_{Q+1}(x)I_{Q-1}(x)}{I_{1}(2x)} \nonumber  \\
   x &\equiv& \frac{1}{2}Vf_{\pi}^{2}M_{\pi}^{2}
\end{eqnarray}
An accurate determination of $f_{\pi}$ in the usual fashion from pseudoscalar-axial vector
correlators is difficult on such small lattices, as the only time window available is
T=1-2 (the axial correlator is antiperiodic and vanishes at T=3). However, it is
apparent from  (\ref{eq:topformula}) that $f_{\pi}$ can be extracted from 
the topological charge
distribution by a one-parameter fit of the dimensionless $x$ variable, 
once the pion masses have
been measured. In Fig. 8 we show the measured distributions (diamonds) of $Q$ for the 
four different sea-quark values as well as the fits to the chiral 
prediction ({\ref{eq:topformula}). The narrowing
of the distribution as one goes to lighter quarks is clearly visible. 
The values of $f_{\pi}$ extracted from the fit $x$ values are given in 
Table \ref{tab:lat}.  At $\kappa=$0.2050 the value of $f_{\pi}$
has been previously extracted from a large ensemble study of axial vector correlators 
using all-point quark
propagators \cite{dpy}: this method gives $f_{\pi}=$0.187$\pm$0.011, close to the 
value 0.21 found from the topological charge fit. Also, the value of $f_{\pi}$ is fairly 
constant over the (limited) range of sea-quark masses studied, as we expect. 
These results certainly contribute to our confidence that
the TDA method builds in all the important low energy chiral physics of QCD.

\begin{figure}
\psfig{figure=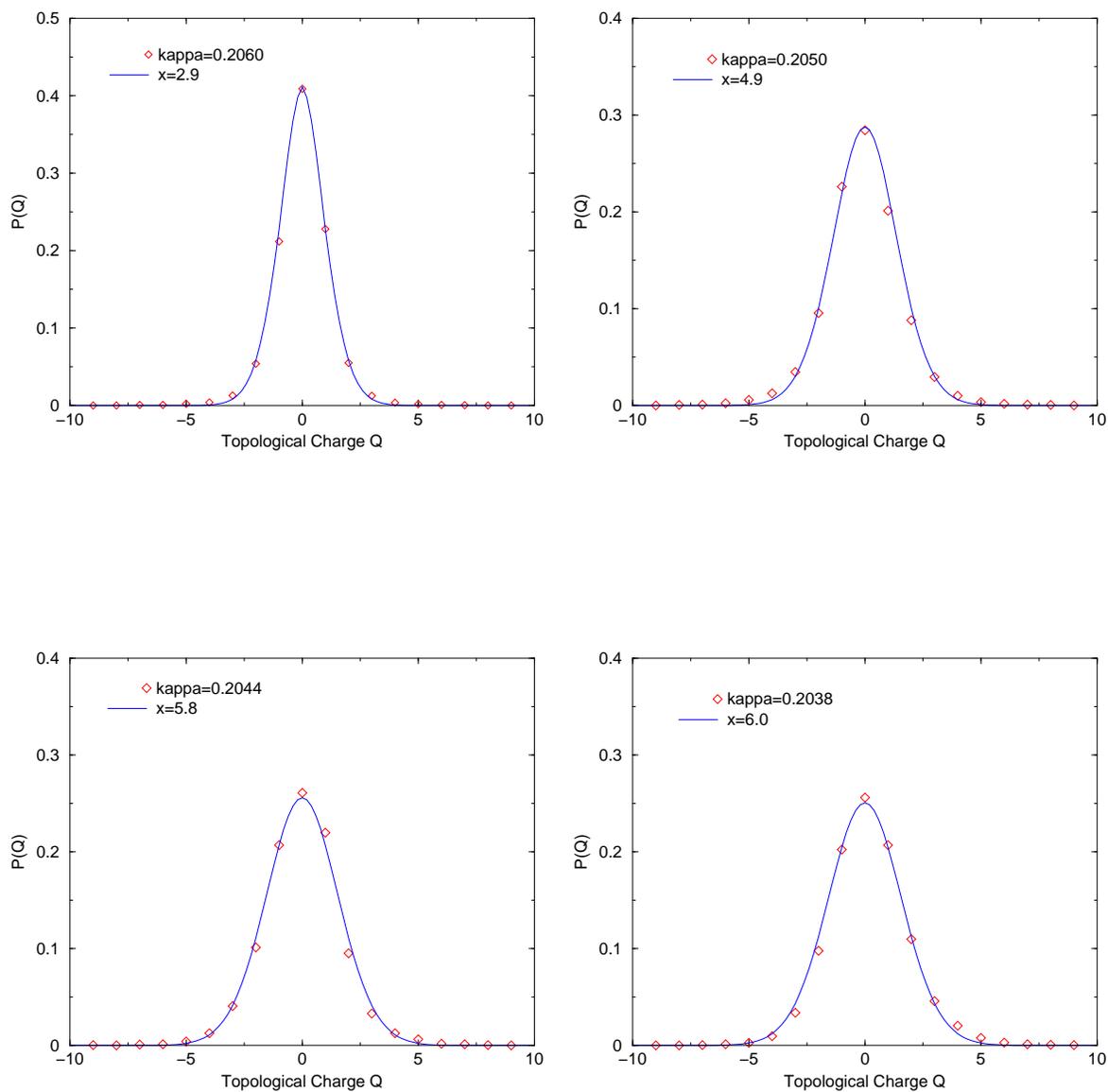,height=2.0\hsize}
\vspace*{-5in}
\caption{Topological charge distributions and fits for $\kappa=$0.2060,0.2050,0.2044,0.2038}
\end{figure}   

\newpage
\section{TDA simulations on finer lattices}

  Although unquenched simulations on physically large but coarse lattices may yield useful
qualitative insights (especially with regard to the dynamics of the simulation process), we
can only expect quantitatively useful results by simulating larger and finer lattices. A
number of TDA simulations on 10$^3$x20 lattices at a lattice spacing $a^{-1}$=1.15 GeV
have therefore been performed to assess the practicality of the TDA method for larger 
lattices (the increase in computational effort required for even larger lattices is discussed
at the end of this section). The gauge action used in this case is a single plaquette one as
improvement is not as important with a lattice spacing of order 0.17 fermi as it was in the 
case of the coarse lattices with spacing 0.4 fermi discussed in the preceding section. However,
we have used clover improvement (with a Sheikoleslami-Wohlert coefficient of $C_{sw}=$1.57)
for the fermions. This allows us to determine the lattice spacing from the rho mass, rather than
the string tension, although the two values are basically quite consistent, as we discuss below.
In the TDA simulations, the lowest 520 eigenvalues were kept, corresponding to a TDA scale
of about 504 MeV. The Lanczos extraction of these eigenvalues for a single configuration takes
about 1.3 hours on a Pentium-4 1.7GHz processor: as this completely dominates the computational
effort, this time also represents a single update step of the TDA simulation for these lattices.

  The preliminary results described in this section were obtained from two separate ensembles
corresponding to runs at $\beta=$5.7 and $\kappa=$0.1420 (h1 run) and 0.1415 (h2 run). To this
point, 100 configurations separated by 50 update steps were obtained for the h2 run and 80
configurations for the lighter h1 run. Most of the discussion will concern the h2 run as finite
volume effects, available statistics and autocorrelation effects are all worse for the h1 run.
The simulations are continuing and we expect to accumulate significantly larger ensembles in the coming 
months, by implementing a completely parallel version of the Lanczos process. However, we
again emphasize here that the acceptance rates for the TDA simulations in the two runs
are essentially identical (0.55 for $\kappa$=0.1415 and 0.53 for $\kappa$=0.1420), once again
illustrating the immunity of the TDA approach to the critical slowing down endemic in HMC approaches
at light quark masses.

\begin{figure}
\psfig{figure=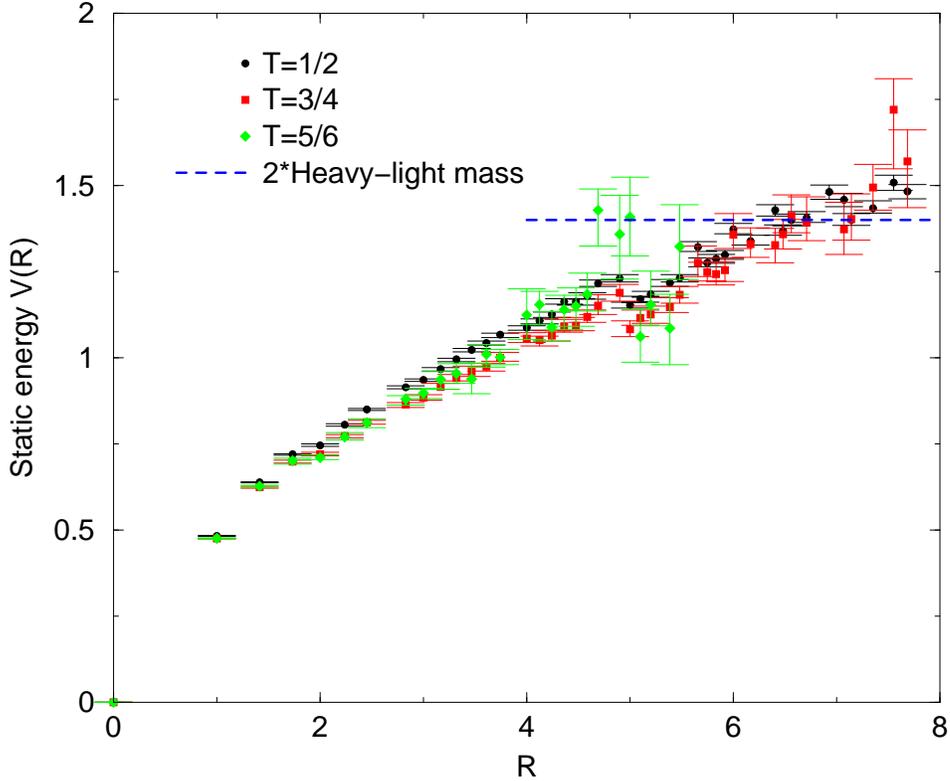,height=0.7\hsize}
\caption{Static Potential for the h2 run, $\kappa=$0.1415}
\end{figure}

  As the spatial extent  of the h2 run lattices is considerably smaller (1.7 fm as opposed to 2.4
 fm for the 6$^4$ lattices) we may expect that the string breaking effects will also be more
difficult to see.  The static potential measured at various Euclidean times is shown in Fig. 9,
and there is as yet  no evidence for a flattening of the potential at distances where the errors
are still reasonable at larger distances (for lattice times $<\simeq$ 6, or about 1 fm). The
expected asymptotic limit (=twice the heavy-light meson mass) is indicated by the dashed line:
with the statistics available, this value is only reached when the errors for Wilson lines
of temporal extent $T>$4 begin to explode. Much
larger statistics will presumably be needed to reach the larger times and distances where stringbreaking
will appear on these lattices.  We can however use the initial rise of the static potential to 
extract a rough lattice scale. Extracting a slope from the region $2\leq R \leq 4$, one finds $a$=0.16 fm.
A more reliable estimate of the scale can be obtained from the rho mass $M_{\rho}=$0.669(30), 
obtained by fitting a set of 200 bootstrapped smeared-local rho propagators, as shown in Fig. 10.
Using the rho mass to fix the lattice scale gives $a^{-1}$=1.15 Gev, $a$=0.17 fm for the h2 run:
as in the case of the 6$^4$ runs discussed in Section 3, the scale is not very sensitive to the sea-quark
mass in this very light regime, and at $\kappa=$0.1420 we find a rho mass of 0.693(30) in lattice units,
giving a lattice scale $a^{-1}$=1.11 GeV.

\begin{figure}
\psfig{figure=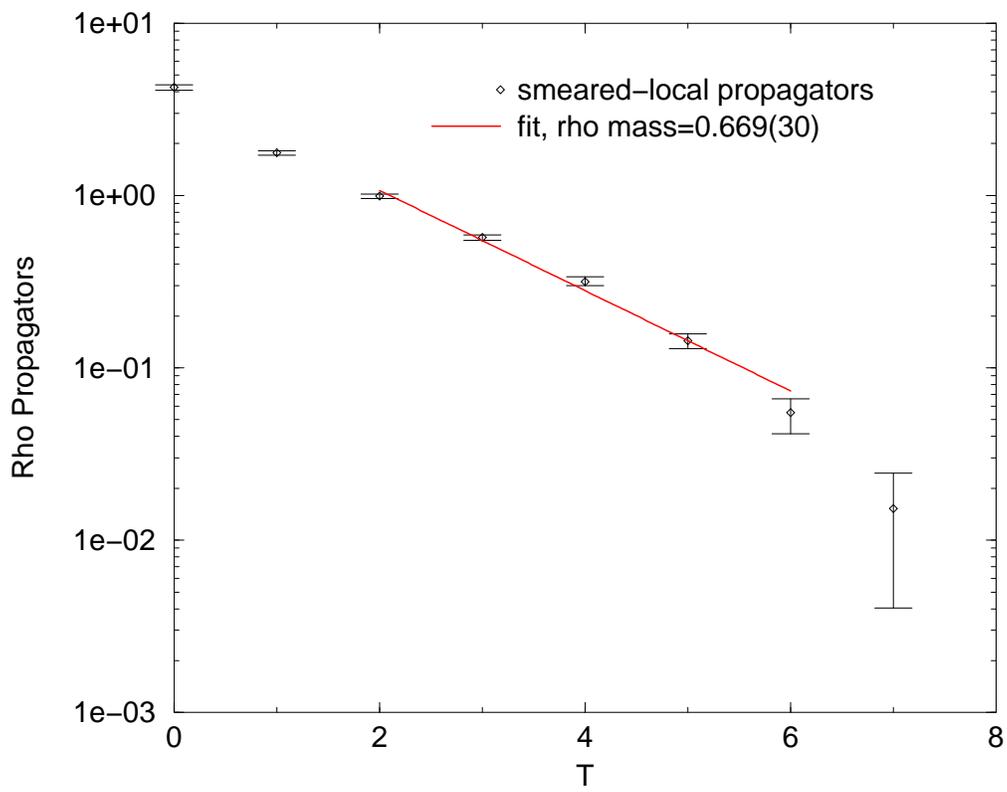,height=0.7\hsize}
\caption{Rho propagator, h2 run, $\kappa=$0.1415}
\end{figure}   

 The pion in the h2 ensemble is already very light and autocorrelations are large, typically
on the order of 100-150 update steps (with the autocorrelation time growing with
Euclidean time) . To analyse the pion propagators, we have therefore 
binned the smeared-local pion propagators from 
99 successive configurations (separated by 50 update steps)
into 33 sets of bin size 3 before generating 66 bootstrap propagators
for a bootstrap analysis. The corresponding average propagator file and fit is shown
in Fig. 11, giving a pion mass of 0.175 in lattice units or 201 MeV. For the
h1 run at $\kappa=$0.1420, the pion is even lighter (Fig. 12), but with the
limited statistics (80 configurations) available so far the errors at larger
Euclidean time are substantial, so an accurate determination of the pion mass
in this case is not possible. The h1 propagator is essentially flat for T$>$4, so
this case is presumably very close to kappa critical. We should point out that
these  h1 propagators were obtained by standard conjugate gradient as the stabilized
biconjugate gradient routines often fail to converge for very light quarks.  Of course,
 on this smaller lattice finite size effects (which were small on the 2.4 fm
lattices for pion masses $> 200$ MeV) may well be significant.

\begin{figure}
\psfig{figure=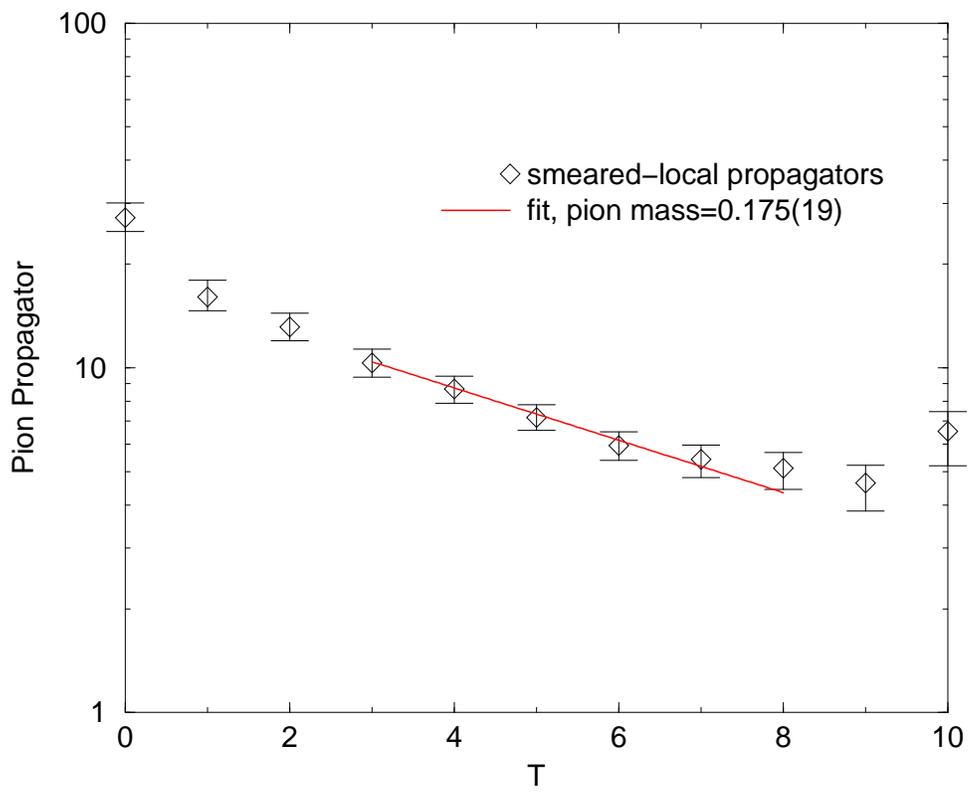,height=0.7\hsize}
\caption{Pion propagator, h2 run, $\kappa=$0.1415}
\end{figure}   

\begin{figure}
\psfig{figure=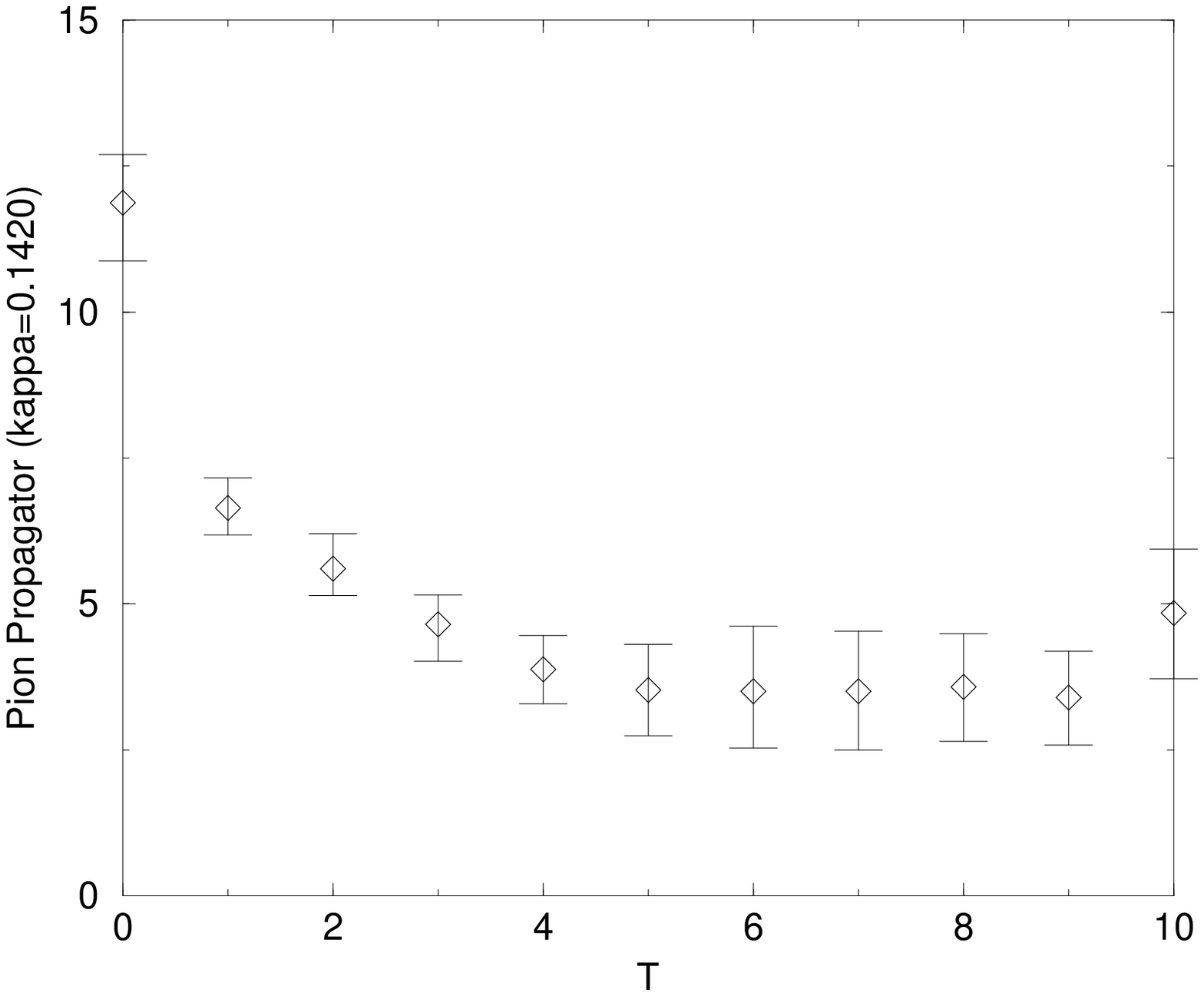,height=0.7\hsize}
\caption{Pion propagator, h1 run, $\kappa=$0.1420}
\end{figure}   

 The anomaly in the quenched theory induced in the scalar isovector channel by the
incomplete cancellation of the quenched eta-prime double pole \cite{scpaper} is by
now well understood. The TDA approximation, while including effects of sea-quark
loops up to fairly high off-shellness {\em exactly}, does not of course treat
valence and sea-quarks identically, so we should expect the appearance of 
incompletely cancelled double pole contributions in isoscalar channels here
also. In the case of the scalar isovector propagator, these contributions
are negative metric and result in the propagator going negative at intermediate
values of Euclidean time. A similar dip is observed in the scalar isovector
propagator obtained form the h2 runs (see Fig. 13), although it is far less
pronounced than in the quenched case. In Section 5 we shall see that the same correlator
is perfectly well-behaved in the TDA+multiboson exact unquenched algorithm, and 
indeed allows a statistically accurate extraction of the eta-prime mass without the need for
subtraction of disconnected contributions.

\begin{figure}
\psfig{figure=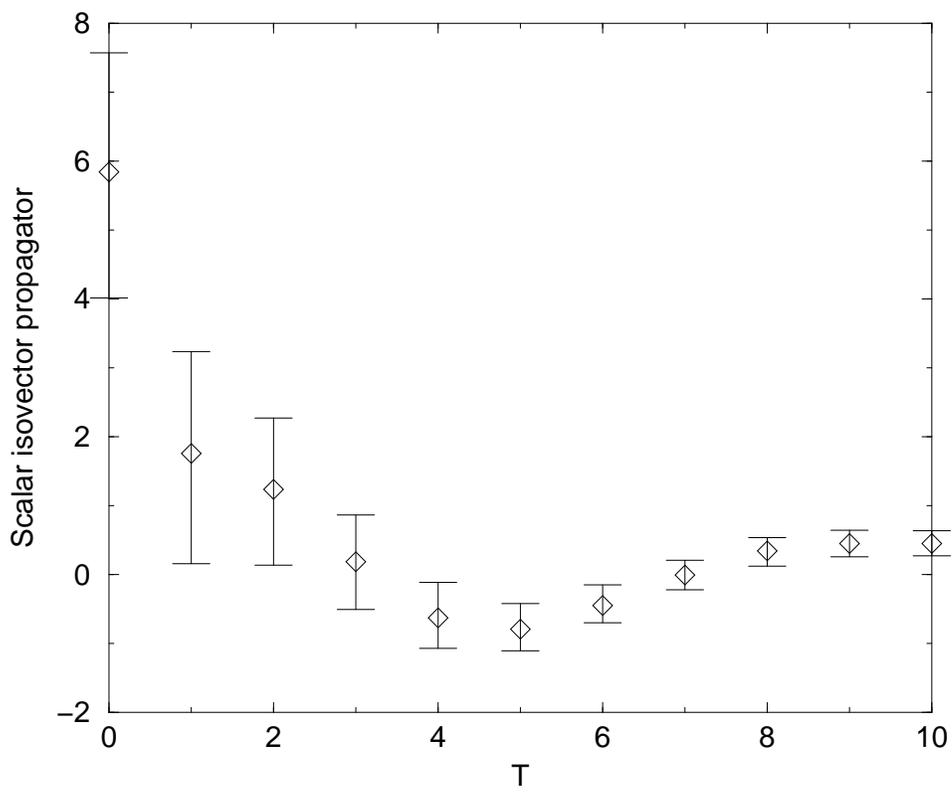,height=0.7\hsize}
\caption{Scalar isovector propagator, h2 run, $\kappa=$0.1415}
\end{figure}   

  The computational effort required to compute all quark eigenmodes up to a fixed $\Lambda_{TDA}$
on lattices of fixed lattice spacing and growing volume $V$ increases like $V^{\alpha}$ where
the exponent $\alpha$ is slightly less than 2. For example, for $\Lambda_{TDA}=$ 450 MeV, we
find that for lattices with $a^{-1}=$1.15 GeV, an update step amounts to approximately 1,3.5 and
35 hours on a Pentium-4 2.2GHz processor on 10$^3$x20 (370 eigenvalues), 12$^3$x24 (850 eigenvalues)
and 16$^3$x32 (2770 eigenvalues) lattices respectively (cf. Table \ref{tab:TDAvol}). 
The scaling properties of the Lanczos
code (employing SSE2 acceleration \cite{SSE2}) on a PC cluster with myrinet interface appear very
good, so a 16 node cluster with 3GHz processors should be adequate for useful simulations (and
comparable in computational effort to the simulations presented here) for even
the physically quite large (2.7 fm)$^3$x(5.4 fm) 16$^3$x32 configurations. 
\begin{table}
\centering
\caption[lattices]{Volume scaling for TDA calculations}
\label{tab:TDAvol}
\begin{tabular}{|c|c|c|c|}
\hline
\hline
 Lattice  & $10^3\times 20$  & $12^3\times 24$ & $16^3\times 32$ \\
\hline
\hline
 time & 1 hrs & 3.5 hrs & 35 hrs \\
\hline
 $N_{\rm eig}$ & 370 & 850 & 2770  \\
\hline
 Lanczos sweeps & 28,000 & 74,000  &  200,000 \\
\hline
\hline
\end{tabular}
\end{table}

In the next section we shall 
see that the inclusion of the ultraviolet part of the determinant by multiboson techniques increases
the computational cost of an update insignificantly, so these estimates hold also for exact 
algorithms where complete  control of the infrared allows probing of the deep chiral limit
without critical slowing down of the MonteCarlo dynamics.

\newpage
\section{Exact Unquenched QCD with light quarks: combining TDA and multiboson methods}

  The evaluation of the ultraviolet contribution to the quark determinant $D_{UV}$ can
be accomplished by the Luescher multiboson technique \cite{Luescher}, as pointed out
previously in \cite{TDA1}. The basic idea of the multiboson technique is to introduce
a series of polynomials in a variable $s$ (shortly to be identified with the square
of the eigenvalues of $H$, for two degenerate sea-quarks) which converge to $s^{-1}$:

\begin{eqnarray}
\lim_{N\rightarrow\infty} P_{N}(s) &=& \frac{1}{s}, \;0\leq s\leq1   \\
  \rm{det}(H^{2}) &=& \lim_{N\rightarrow\infty} (\rm{det}P_{N}(H^{2}))^{-1}  \\
  P_{N}(H^{2}) &\simeq& \prod_{k=1}^{N} ((H-\mu_{k})^{2}+\nu_{k}^{2}) \\
  S_{\rm bosonic} &=& \sum_{k=1}^{N}\sum_{x} \{|(H-\mu_{k})\phi_{k}(x)|^{2}+\nu_{k}^{2}|\phi_{k}(x)|^{2}\}
\end{eqnarray}
Specifically, it is convenient to pick Chebyshev polynomials so that with
$u=(s-\epsilon)/(1-\epsilon)$ and $\cos{\theta}=2u-1$, $T^{*}_{r}(u)=\cos{(ru)}$. Then
\begin{equation}
  P_{N}(s) \equiv (1+\rho T^{*}_{N+1}(u))/s
\end{equation}
with $\rho$ chosen so that $P_{N}(s)$ has a finite limit as $s\rightarrow 0$. With these
choices, $sP_{N}(s)$ differs from unity in the interval $\epsilon \leq s \leq 1$ by an
amount less than $2(\frac{1-\sqrt{\epsilon}}{1+\sqrt{\epsilon}})^{N+1}$.

 The $N$ roots of the polynomial $P_{N}(s)$ typically lie on an ellipse in the complex plane
surrounding the spectrum of $H^2$ (with $H$ rescaled so that the spectrum of $H^2$
lies between 0 and 1). An example, for values of $N$=20,80, is shown in Fig. 8. The
essential point is that accurate control of the infrared spectral region requires the
number of multiboson fields  $N$ to be chosen large (specifically, if we demand a
fixed relative error uniformly in the range $\epsilon \leq s \leq 1$, then we must
 hold $\sqrt{\epsilon}N$ fixed as $\epsilon\rightarrow 0$). This then forces many
 of the bosonic fields to appear in the action with small ``masses" $\nu_{k}$,
 which in turn leads to critical slowing down in the multiboson sector.

\begin{figure}
\psfig{figure=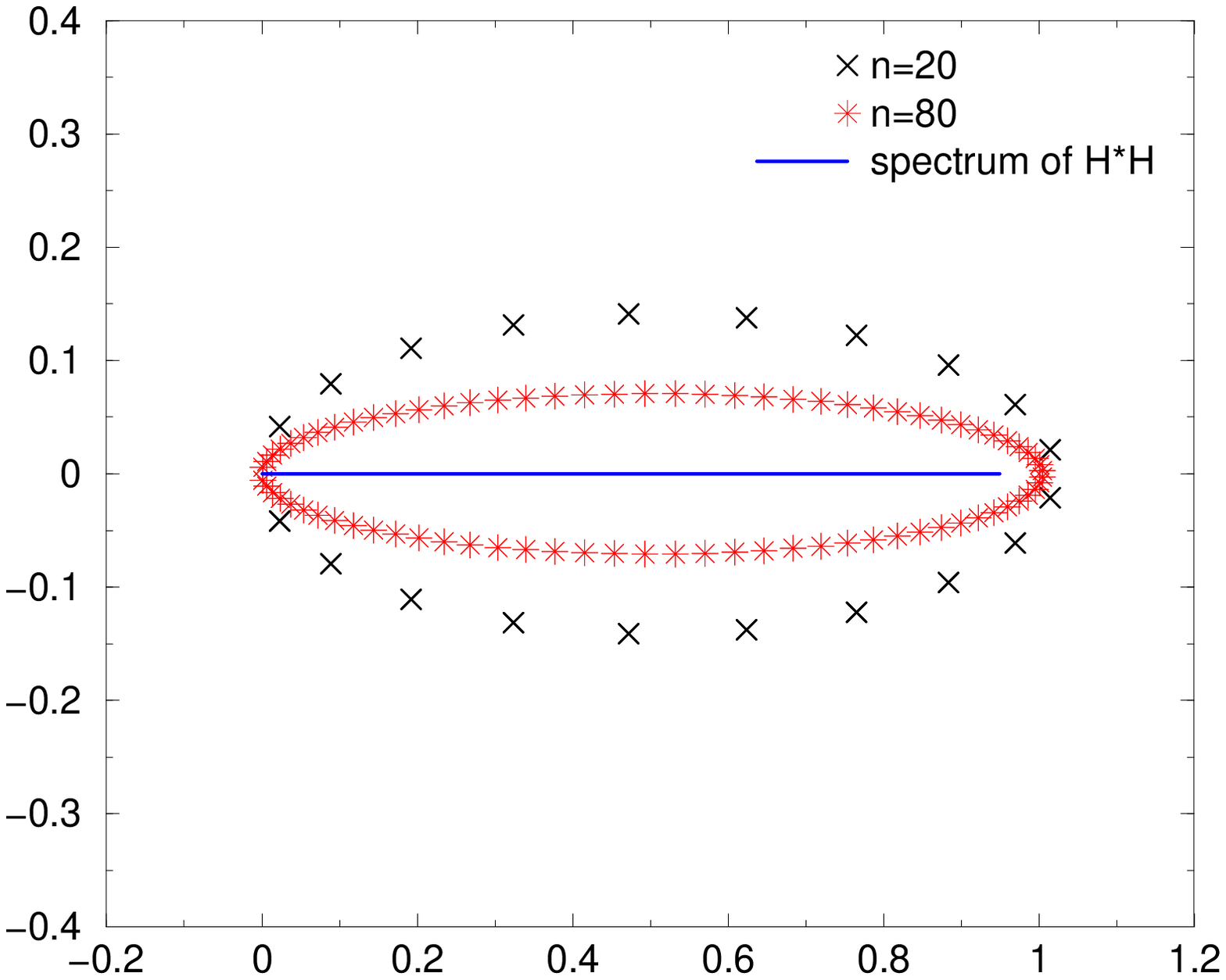,height=0.7\hsize}
\caption{Zeroes of the approximating multiboson polynomial for 20,80 boson fields}
\end{figure}

 Evidently, the exact control over a substantial segment of the infrared quark spectrum
provided in the TDA approach suggests that we should be able to reduce substantially the
number of multiboson fields, with a corresponding amelioration of the critical slowing
down problem. Define a determinantal compensation factor for $N$ multiboson fields as
follows
\begin{equation}
    D_{CF}(N,N_{\rm eig}) \equiv \ln{(\prod_{i=1}^{N_{\rm eig}}\lambda_{i}^{2}P_{N}(\lambda_{i}^{2}))}
\end{equation}
where $N_{\rm eig}$ is the number of eigenvalues $\lambda_{i}$ (ordered in absolute value) of $H$
calculated in the TDA approach. We expect that $D_{CF}$ should converge to a well-defined limit
once $\lambda_{N_{\rm eig}}^{2} >> \epsilon$. At this point, the compensation factor $D_{CF}$
can be used in an accept/reject step to correct the approximate determinant generated by the 
multiboson part of the action. To get a quantitative feeling for how rapidly this convergence 
takes effect, we show two examples in Figs 15,16.   In Fig. 15 we consider pairs of adjacent
 configurations in a simulation in which the 1000 lowest eigenvalues of $H^{2}$ on a $6^4$ lattice (with
lattice spacing $a\simeq$0.4F)
 are exactly
computed by the Lanczos method, while the multiboson action corresponds to $N=20,\epsilon=0.02$.
 The plot shows the difference of $D_{CF}$ for two successive configurations (needed for the
accept/reject step) as a function of the number of eigenvalues $N_{\rm eig}$ included in the
 computation of $D_{CF}$. Evidently, the small number of pseudofermion fields used means that
the simulated determinant is very inaccurate until several hundred exactly computed eigenvalues
are included, at which point the needed determinantal compensation factor converges rapidly.
The dependence on $N_{\rm eig}$ is shown for 6 separate pairs of adjacent configurations in
the MonteCarlo sequence. In Fig. 16 we show a similar plot for 10$^3$x20 lattices (lattice spacing
$a\simeq 0.17$ fm) up to a maximum of $N_{\rm eig}$=600, with $N=50,\epsilon=0.003$.
\begin{figure}
\psfig{figure=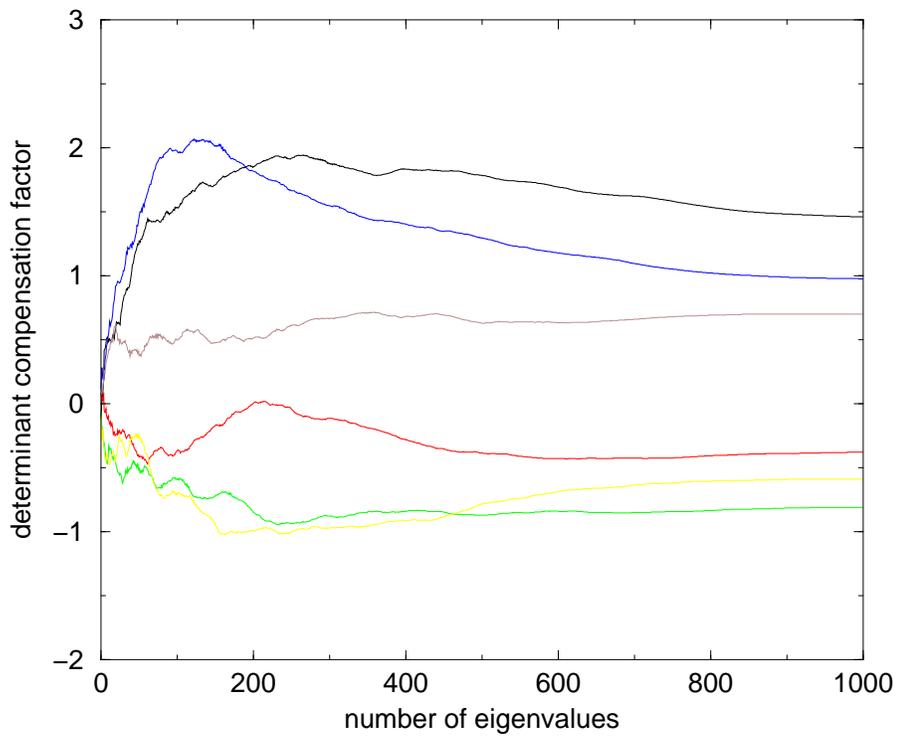,height=0.65\hsize}
\caption{Convergence of determinant correction, 6$^4$ lattice ($N=$20,$\epsilon=$0.02)} 
\end{figure} 
\begin{figure}
\psfig{figure=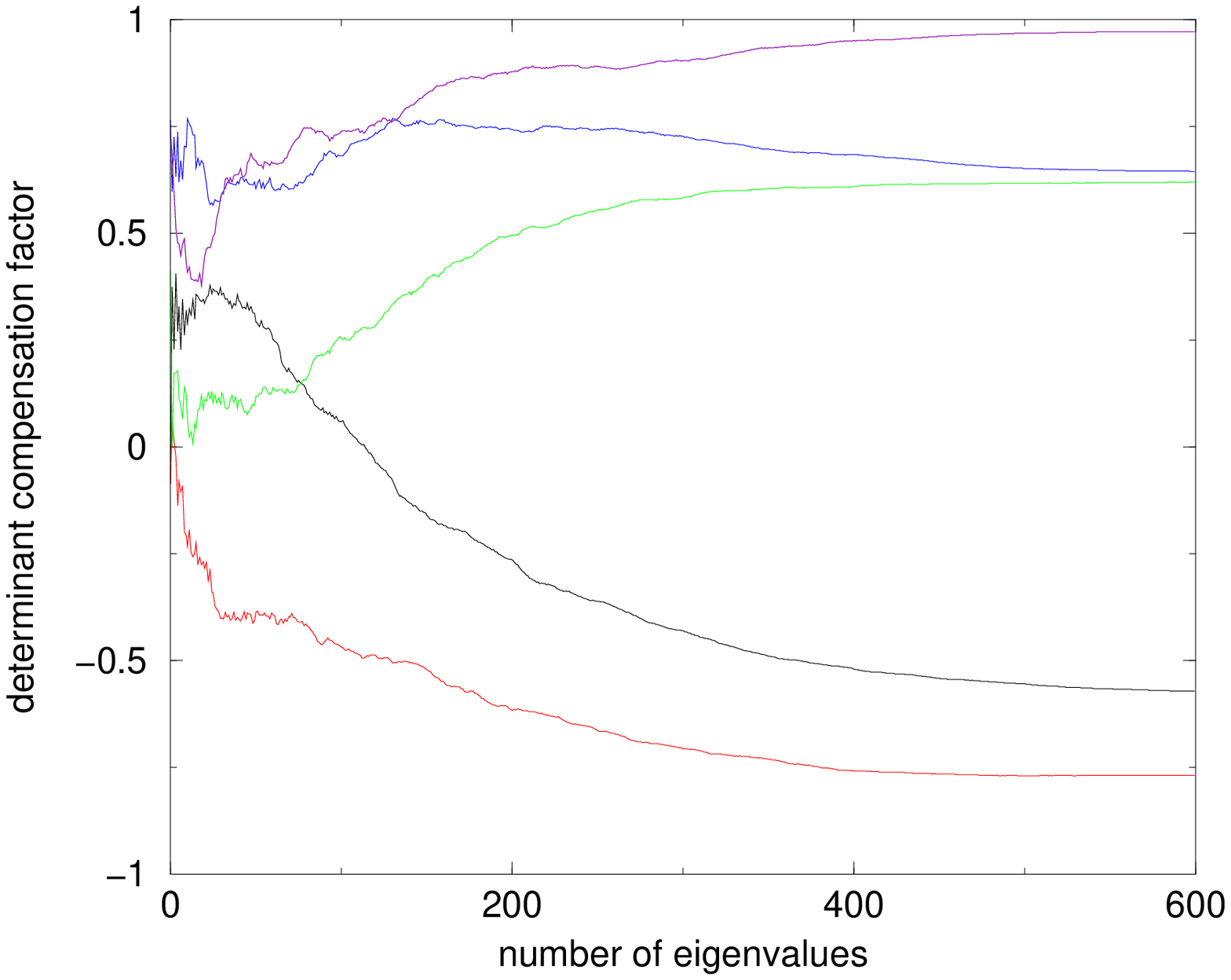,height=0.65\hsize}
\caption{Convergence of determinant correction, 10$^3$x20 lattice ($N=$50,$\epsilon=$0.003)} 
\end{figure}

In order to get a feeling for the basic features of the MonteCarlo
dynamics of the combined TDA and multiboson approach described here,
we have performed simulations on large, coarse lattices, roughly similar
to the ensembles described in Section 3. The gauge action was improved
exactly as described there, but with the couplings $\beta_{\rm plaq}$=3.65,
$\beta_{\rm trt}$=0.75. The lowest 1000 eigenvalues (corresponding to 
a TDA scale of about 560 MeV) were computed exactly, and  20 multiboson
fields, with $\epsilon=$0.02 were used to compute the UV part of the 
determinant. The full algorithm breaks into the following steps
\begin{enumerate}
\item Multiboson and gauge updates (satisfying detailed balance):\\
(a) 1 overrelaxation update of the multiboson fields (with overrelaxation
parameter $\omega$=1.9)\\
(b) gauge fields update consisting of 10 hit Metropolis, an overrelaxation
step and another 10 hit Metropolis update\\
(c) 1 overrelaxation update of the multiboson fields
\item Computation of the determinant compensation factor $D_{CF}$, which is then
used in an accept/reject step for the new gauge and multiboson fields. The
acceptance rate for this step is similar to that in the pure TDA approach,
showing little sensitivity to the quark mass in the chiral region 
(e.g.  ranging from 46\% at $\kappa=0.1915$, corresponding to a pion mass of 380 MeV
to 36\%  at $\kappa=$0.1930, corresponding to a pion
mass of 230 MeV).
\end{enumerate}
 At $\kappa$=0.1920, where we have already equilibrated and generated a reasonably
large ensemble,  we find a lattice scale of $a = 0.36$ fm (from string tension
measurements) or $a^{-1}\simeq$ 550 MeV. Thus these fully unquenched coarse
lattices are roughly similar in size to the ones described in Section 3.

  As a simple example of a fully unquenched quantity computed in this approach,
where consistent inclusion of all quark eigenmodes is important in avoiding
unphysical anomalies, we show in Fig. 17 the pseudoscalar (pion) and scalar 
isovector propagators obtained from 100 configurations at $\kappa$=0.1920.
One obtains a pion mass of 0.60 in lattice units, while the exponential 
fall of the scalar correlator (corresponding to an intermediate s-wave two-body state
of a pion and an $\eta^{\prime}$) corresponds to a mass of 1.94. Subtracting these
we find an $\eta^{\prime}$ mass of 1.34 in lattice units, or about 735 MeV, not far
from the value of 715 MeV expected in a fictional two-light flavor world \cite{schilling}.
The advantage of this approach to the $\eta^{\prime}$ mass is that the need to
subtract disconnected diagrams is completely circumvented! Simulations of this exactly
unquenched two-flavor system at lighter quark masses  (down to the physical up and
down quark masses) are continuing to check the chiral
extrapolation of this result, and we are also beginning simulations with 2+1 sea-quark flavors
(up, down and strange dynamical quarks) in order to study the $\eta-\eta^{\prime}$
spectrum in a more realistic setting.

\begin{figure}
\psfig{figure=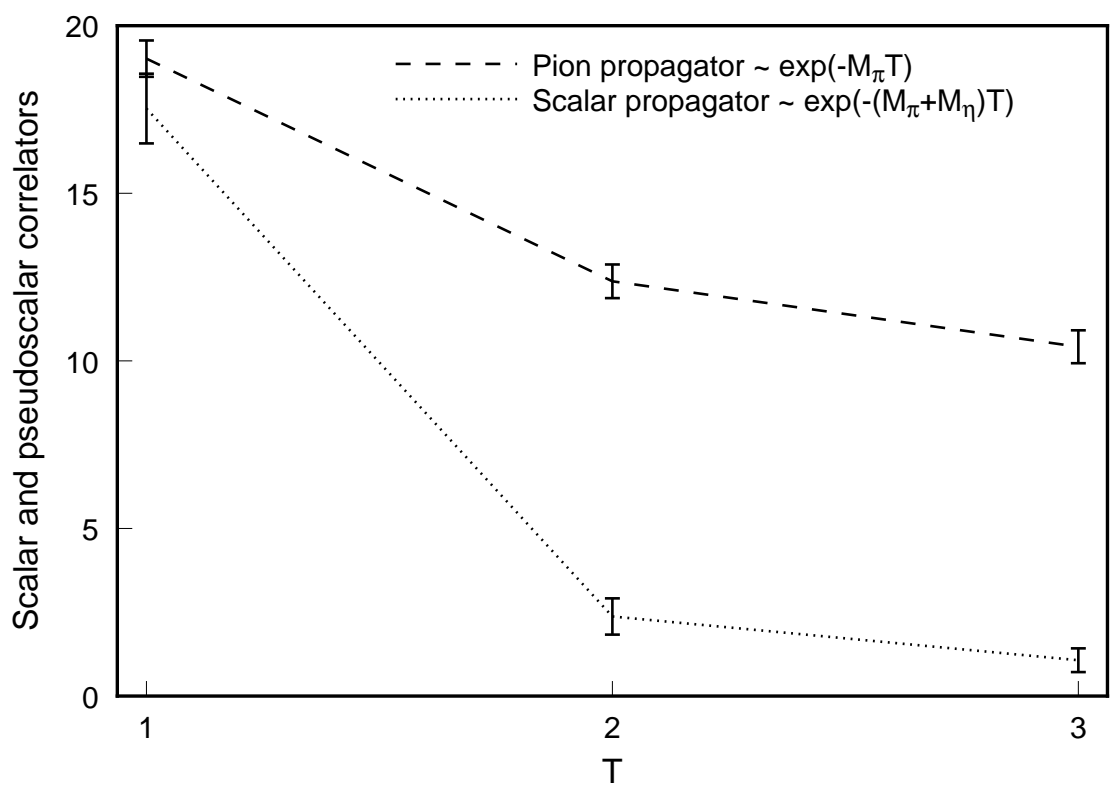,height=0.7\hsize}
\caption{Scalar and Pseudoscalar Isovector propagators, fully unquenched}
\end{figure}

\newpage
\section{Acknowledgements}
The authors are grateful for useful conversations with M. diPierro and
H. Thacker. 
The work of A. Duncan  was supported in part by 
NSF grant PHY00-88946. The work of E. Eichten was performed
at the Fermi National Accelerator Laboratory, which is operated by 
University Research Association, Inc., under contract DE-AC02-76CD03000.

\end{document}